\documentclass[aps
, prl
, twocolumn
, nofootinbib 
, superscriptaddress
]{revtex4-2}

\usepackage{graphicx}
\usepackage{xcolor}
\usepackage{tikz}
\usepackage[caption=false]{subfig}
\usepackage{mathrsfs,mathtools}
\usepackage{physics,amssymb}
\usepackage{bm}
\usepackage{braket}
\usepackage{listings}
\usepackage{cases}
\usepackage{comment}
\usepackage{soul}
\usepackage{cancel}
\usepackage{cases}
\usepackage[utf8]{inputenc}
\usepackage{url}
\usepackage{longtable}
\usepackage[normalem]{ulem}
\usepackage{xspace}
\usepackage[colorlinks=true
,urlcolor=DARKBLUE
,anchorcolor=DARKBLUE
,citecolor=DARKBLUE
,filecolor=DARKBLUE
,linkcolor=DARKBLUE
,menucolor=DARKBLUE
,linktocpage=true
,pdfproducer=medialab
,pdfa=true
]{hyperref}

\newcommand{\Del}{\partial}

\newcommand{\ee}{\mathrm{e}}
\newcommand{\diag}{\mathrm{diag}}
\newcommand{\sign}{\mathrm{sign}}
\newcommand{\Mpl}{M_\mathrm{Pl}}
\newcommand{\ns}{n_{\mathrm{s}}}
\newcommand{\cs}{c_{\mathrm{s}}}
\newcommand{\eq}{\mathrm{eq}}
\newcommand{\GW}{\mathrm{GW}}
\newcommand{\SO}{\mathrm{SO}}

\newcommand{\G}{\Gamma}

\newcommand{\uf}{\mathrm{f}}

\newcommand{\calL}{\mathcal{L}}

\newcommand{\um}{\mathrm{m}}

\newcommand{\calO}{\mathcal{O}}

\newcommand{\calP}{\mathcal{P}}

\newcommand{\uR}{\mathrm{R}}
\newcommand{\ur}{\mathrm{r}}
\newcommand{\us}{\mathrm{s}}

\newcommand{\VP}{\varphi}

\newcommand{\bae}[1]{\begin{align} #1 \end{align}}
\newcommand{\bce}[1]{\begin{cases} #1 \end{cases}}
\newcommand{\dps}{\displaystyle}

\definecolor{MONZA}{HTML}{CF000F}
\definecolor{DARKBLUE}{HTML}{00008b}
\definecolor{DARKMAGENTA}{HTML}{8b008b}

\begin{document}
\title{\boldmath Minimal $k$-inflation in light of the conformal metric-affine geometry
}
\date{\today}

\author{Yusuke Mikura}
\email{mikura.yusuke@e.mbox.nagoya-u.ac.jp}
\affiliation{Department of Physics, Nagoya University, Nagoya 464-8602, Japan}
\author{Yuichiro Tada}
\email{tada.yuichiro@e.mbox.nagoya-u.ac.jp}
\affiliation{Department of Physics, Nagoya University, Nagoya 464-8602, Japan}
\author{Shuichiro Yokoyama} \email{shu@kmi.nagoya-u.ac.jp}
\affiliation{%
  Kobayashi Maskawa Institute, Nagoya University, Chikusa, Aichi
  464-8602, Japan}%
\affiliation{%
  Kavli IPMU (WPI), UTIAS, The University of Tokyo, Kashiwa,
  Chiba 277-8583, Japan}%

\begin{abstract}
We motivate a minimal realization of slow-roll $k$-inflation by incorporating the local conformal symmetry and the broken global $\SO(1,1)$ symmetry in the metric-affine geometry. With use of the metric-affine geometry where both the metric and the affine connection are treated as independent variables, the local conformal symmetry can be preserved in each term of the Lagrangian and thus higher derivatives of scalar fields can be easily added in a conformally invariant way. Predictions of this minimal slow-roll $k$-inflation, $\ns\sim 0.96$, $r\sim 0.005$, and $\cs~\sim 0.03$, are not only consistent with current observational data but also have a prospect to be tested by forthcoming observations.
\end{abstract}

\maketitle

\paragraph{Introduction.---\hspace{-0.8em}}

Nowadays cosmic inflation is considered as the standard paradigm for the early universe~\cite{Starobinsky:1980te,Sato:1980yn,Guth:1980zm,Linde:1981mu,Albrecht:1982wi,Linde:1983gd}. Since the advent of its concept, a considerable number of inflationary models have been proposed for its realization from different motivations~\cite{Lyth:1998xn}. One typical scenario is that the accelerated expansion is driven by the potential energy of a scalar field, called inflaton. In particular, a class of $\alpha$-attractors attracts more and more attention of cosmologists because its predictions are in great agreement with observations of the cosmic microwave background~(CMB) represented by the Planck Collaboration~\cite{Akrami:2018odb}. Besides its phenomenological success, an intriguing aspect lies in its guiding principles, the \emph{local conformal symmetry} and \emph{slightly broken global symmetry}.
The former symmetry is an important concept in many physical contexts (see, e.g., Ref.~\cite{Bars:2013yba}) and the latter one can be powerful to control fields' interactions. Indeed, successful representatives of the class, the T-model~\cite{Kallosh:2013hoa,Kallosh:2013lkr,Kallosh:2013yoa,Kallosh:2014rga,Linde:2015uga} and the Starobinsky model~\cite{Starobinsky:1980te}, can be derived from the viewpoint of the local conformal symmetry and the broken global $\SO(1,1)$ symmetry, where the inflaton can be recognized as the pseudo Nambu-Goldstone mode of the broken global symmetry.

Inflationary models in light of the local conformal symmetry, dubbed 
\emph{conformal inflation}, have been investigated mainly in the (pseudo-)Riemannian spacetime geometry, (see, e.g., Refs.~\cite{Kallosh:2013hoa,Kallosh:2013pby,Kallosh:2013daa,Ferrara:2010in} for a first concept and Refs.~\cite{Buchmuller:2012ex,Buchmuller:2013zfa,Brax:2014baa,Ishiwata:2018dxg,Bernardo:2017xcm} for its phenomenology), where the metric $g$ is treated as an independent variable and the affine connection $\G$ is determined only by the metric, i.e., it is usually the Levi-Civita connection. In this geometry, the Ricci scalar exhibits a non-trivial change as
\bae{
    R(g)\to\tilde{R}(g)=\ee^{2\sigma(x)}\pqty{R(g)-6\ee^{\sigma(x)}\Box\ee^{-\sigma(x)}}, \label{R-trans-GR}
}
under the local conformal transformation $g_{\mu\nu}\to\tilde{g}_{\mu\nu}=\ee^{-2\sigma(x)}g_{\mu\nu}$ in the four dimensional spacetime. Consequently, with a scalar field transforming as $S\to\tilde{S}=\ee^{\sigma(x)}S$, the form of a locally conformal-invariant action is restricted as
\bae{
    \calL
    \!\supset\!\sqrt{-g}\pqty{\!\frac{1}{12}S^2R(g)\!-\!X(S)\!} \qc
    X(S)\!\coloneqq\! -\frac{1}{2}g^{\mu\nu}\Del_\mu S\Del_\nu S,
    \label{eq:Lag_conformal}
}
where a sign in front of the kinetic term $X(S)$ needs to be negative to cancel out extra $\sigma$-derivatives in Eq.~(\Ref{R-trans-GR}) by the scalar kinetic term.
The scalar $S$ thus behaves as a ghost in this case and should not appear in the physical spectrum.

From a geometrical viewpoint, one can use a generalized geometry called the \emph{metric-affine} one in which the affine connection is treated a priori as an independent variable along with the metric (see, e.g., Refs.~\cite{Hehl:1994ue,Kleyn:2004yj,Vitagliano:2010sr, Vitagliano:2013rna,Vazirian:2013baa,Aoki:2018lwx} and references therein).
Applications of this geometry (or so-called Palatini formalism) to cosmology have gained increasing attention and a decent number of topics have been discussed in, e.g., Refs.~\cite{Bauer:2010jg,Bauer:2008zj,Shimada:2018lnm,Takahashi:2020car,Sotiriou:2006qn,Sotiriou:2008rp,Sotiriou:2009xt,Jarv:2017azx,Helpin:2019kcq,Kubota:2020ehu,Enckell:2018kkc,Rasanen:2018ihz,Bauer:2010jg,Tamanini:2010uq,Rasanen:2017ivk,Tenkanen:2017jih,Racioppi:2017spw,Markkanen:2017tun,Jarv:2017azx,Racioppi:2018zoy,Carrilho:2018ffi,Enckell:2018hmo,Antoniadis:2018ywb,Rasanen:2018fom,Kannike:2018zwn,Almeida:2018oid,Antoniadis:2018yfq,Azri:2017uor,Edery:2019txq,Azri:2019ffj,Gialamas:2019nly,Karam:2021wzz,Gialamas:2020vto,Gialamas:2020snr,Dimopoulos:2020pas} (see also Ref.~\cite{Tenkanen:2020dge} for a recent review).
In this geometry, the Riemann tensor is a function only of the connection, being free from the metric, and thus can be covariant under the local conformal transformation.
Furthermore, the naturally introduced vector, the non-metricity $Q_\mu\coloneqq-g^{\alpha\beta}\nabla_\mu g_{\alpha\beta}$, plays the role of the conformal gauge field, which implements the covariant derivative $D_\mu\coloneqq\partial_\mu-\frac{1}{8}Q_\mu$ for scalars.\footnote{The non-metricity is non-dynamical in our case. Inflation with the dynamical non-metricity known as the \emph{Weyl gauge field} is discussed in Refs.~\cite{Ferrara:2010in,Ferreira:2019zzx,Bars:2013yba,Tang:2018mhn,Barnaveli:2018dxo,Tang:2019olx,Tang:2019uex,Ghilencea:2018thl,Ghilencea:2020piz,Ghilencea:2020rxc,Ghilencea:2019rqj}.}
Consequently, the conformal invariance can be preserved in each term of the Lagrangian in this geometry~\cite{2019Univ....5...82I}.
This independency is beneficial in 
various ways: e.g., it can be compatible with many kinds of extra symmetries and generalize the conformal class of potential-driven inflation as we discussed in the previous letter~\cite{Mikura:2020qhc}. On the other kinetic side, the metric-affine geometry allows higher derivatives to be solely included in the Lagrangian in a conformally invariant way, 
being free from a specific structure~\eqref{eq:Lag_conformal} required in the Riemannian case.

It is widely known that such non-standard kinetic terms can bring about inflation even if the potential term is absent, which we call kinetically driven inflation as a large class. The concept originates in $k$-inflation proposed in Refs.~\cite{ArmendarizPicon:1999rj,Garriga:1999vw} and many models in this class such as the Dirac-Born-Infeld (DBI) model~\cite{Silverstein:2003hf}, the dilatonic ghost condensate~\cite{ArkaniHamed:2003uy,ArkaniHamed:2003uz}, and 
G-inflation~\cite{Kobayashi:2010cm,Kobayashi:2011nu} have been later studied.

In this Letter, we motivate a minimal realization of slow-roll $k$-inflation by the local conformal symmetry and slightly broken global symmetry (specifically, $\SO(1,1)$) in the metric-affine geometry.
We show that predictions of our minimal setup, the spectral index $\ns \sim 0.96$, the tensor-to-scalar ratio $r \sim 0.005$, and the sound speed $\cs \sim 0.03$, are not only consistent with current observational data but also testable by future cosmological observations such as CMB, large-scale structures, and also stochastic gravitational waves.
We adopt the natural unit $c=\hbar=1$ and the sign of the Minkowski metric is defined by $\eta_{\mu\nu}=\diag(-1,1,1,1)$ throughout.

\paragraph{Conformal metric-affine geometry and its compatibility with higher derivatives.---\hspace{-0.8em}}

The metric-affine geometry treats both the metric and the affine connection as independent variables. In this geometry, the local conformal transformation is defined by the change of the metric, while the affine connection is left unaffected in contrast to the Riemannian one:
\bae{
    g_{\mu\nu}\to\tilde{g}_{\mu\nu}=\ee^{-2\sigma(x)}g_{\mu\nu} \qc
    \Gamma^\rho_{\mu\nu}\to\tilde{\Gamma}^\rho_{\mu\nu}=\Gamma^\rho_{\mu\nu}.
}
Since the Ricci tensor $R_{\mu\nu}$ is a function only of the affine connection, it is obvious that the Ricci scalar transforms covariantly under the conformal transformation as
\bae{
    R(g,\Gamma)=g^{\mu\nu}R_{\mu\nu}(\Gamma)\to R(\tilde{g},\tilde{\Gamma})=\ee^{2\sigma(x)}R(g,\Gamma).
    }
This covariant feature of the Ricci scalar points out that the non-minimal coupling term $\sqrt{-g}S^2R(g,\G)$ is conformally invariant by itself without any help of the scalar kinetic term. 
The conformal invariance of the scalar kinetic term can be also accomplished with the help of the naturally introduced vector field, i.e., the non-metricity~\cite{Kleyn:2004yj}
\bae{
    Q_\mu=g^{\alpha\beta}Q_{\mu\alpha\beta}\coloneqq-g^{\alpha\beta}\nabla_\mu g_{\alpha\beta}.
}
Because it behaves as a gauge field associated with the conformal transformation,
\bae{
    Q_\mu\to\tilde{Q}_\mu=Q_\mu+8\partial_\mu\sigma,
}
the covariant derivative for scalar fields can be defined as~\cite{2019Univ....5...82I}
\bae{
    D_\mu S\coloneqq\pqty{\partial_\mu-\frac{1}{8}Q_\mu}S,
}
and one can easily show that the scalar kinetic term defined by $D_\mu$ transforms covariantly under the conformal transformation:
\bae{
    \hat{X}(S)\coloneqq-\frac{1}{2}g^{\mu\nu}D_\mu S D_\nu S\to \ee^{4\sigma(x)}\hat{X}(S).
}

This independency makes it possible to contain higher derivatives in the Lagrangian easily in a conformal-invariant way. A higher derivative term can be kept invariant if it is properly divided by the scalar as
\bae{
    \calL\supset\sqrt{-g}\left[\frac{C_n}{S^{4(n-1)}}\hat{X}^n\right],
}
where $C_n$ $(n=1,2,\cdots)$ is a dimensionless coupling constant.

Although higher derivatives can be easily compatible with the local conformal symmetry in the metric-affine geometry, such a scalar cannot be employed yet for, e.g., the inflation mechanism because it is removed from the theory by fixing the gauge symmetry. Hence one often introduces another scalar which plays a role in exhibiting the inflaton's dynamical degree of freedom.
Upon adding another scalar, we can control their interactions with use of some global symmetry groups.
For example, an $\SO(1,1)$ symmetry~\cite{Kallosh:2013hoa} between two scalars $\chi$ and $\phi$ allows us the following conformal Lagrangian up to $\hat{X}^2$-terms:\footnote{The potential term $\frac{\lambda}{4}(\chi^2-\phi^2)^2$ investigated in our previous work~\cite{Mikura:2020qhc} is also allowed by this symmetry, but we neglect it for simplicity in this work.}
\bae{
    \calL=\sqrt{-g}\left[\frac{\chi^2-\phi^2}{12}R(g,\Gamma)+\beta 
    \pqty{\hat{X}(\chi)-\hat{X}(\phi)} \right. \nonumber \\ \left.+\gamma^4\frac{(\hat{X}(\chi)-\hat{X}(\phi))^2}{(\chi^2-\phi^2)^2}\right],
}
where $\beta$ and $\gamma$ are positive and dimensionless coupling constants. Thanks to the conformal symmetry, two scalars can be unified into one canonically normalized scalar $\VP$ by gauge fixing. As a simple choice, it is convenient to take the so-called \emph{rapidity gauge}~\cite{Kallosh:2013hoa,Kallosh:2013yoa} defined through
\bae{
    \!\chi\!=\!\sqrt{6}\Mpl\,
    \mathrm{cosh}\frac{\VP}{\sqrt{6}\Mpl} \qc 
    \phi\!=\!\sqrt{6}\Mpl\,\mathrm{sinh}\frac{\VP}{\sqrt{6}\Mpl}.
}
$\Mpl$ is the reduced Planck mass. 
In this gauge, the non-minimal coupling term is fixed as the Einstein-Hilbert term and the Lagrangian is reduced to
\bae{
    \calL&=\frac{\Mpl^2}{2}R-\beta
    X(\VP)-\frac{3\Mpl^2}{64}\beta
    Q^\mu Q_\mu \nonumber
    \\
    &\qquad+\frac{\gamma^4}{36\Mpl^4}\left(X^2(\VP)+\frac{3\Mpl^2}{32}Q^\mu Q_\mu X(\VP) \right. \nonumber \\
    &\qquad\qquad\left.+\frac{9\Mpl^4}{64\cdot 64}(Q^\mu Q_\mu)^2\right),
}
where $X(\VP)=-\Del_\mu \VP \Del^\mu \VP/2$ is the kinetic term constructed with a usual derivative. We note that the Ricci scalar is free from the non-metricity $Q_\mu$ (see, e.g., Refs.~\cite{Sotiriou:2006qn,2019Univ....5...82I}), and then its stationary solutions are determined only by explicit $Q_\mu$-terms in the action.
One trivial solution is found as $Q_\mu=0$,\footnote{The other non-trivial solution is $Q^\mu Q_\mu=\frac{64}{\gamma^4}\left(6\Mpl^2\beta-\frac{\gamma^4}{3\Mpl^2}X(\VP)\right)$ and leads to $\calL=\frac{\Mpl^2}{2}R-9\Mpl^4\frac{\beta^2}{\gamma^4}$. We leave this non-trivial branch for a future issue.
At least the attractor stability of the trivial branch~\eqref{Lagrangian0} is confirmed~\cite{ArmendarizPicon:1999rj,Garriga:1999vw}.} which leads to
\bae{
    \calL
    =\frac{\Mpl^2}{2}R-\beta
    X(\VP)+\frac{\gamma^4}{36\Mpl^4}X^2(\VP). \label{Lagrangian0}
}
This is nothing but a minimal realization of the so-called \emph{$k$-inflation}~\cite{ArmendarizPicon:1999rj,Garriga:1999vw}.
The inflaton $\varphi$ realizes the kinetically driven de Sitter universe with the stationary momentum of $P(X)=-\beta X+\frac{\gamma^4}{36\Mpl^4}X^2$, i.e., $X=\dot{\varphi}^2/2=\frac{18\beta\Mpl^4}{\gamma^4}$ at the background level.

Inflation cannot end if both $\beta$ and $\gamma$ are constant.
Then we explicitly break the $\SO(1,1)$ symmetry by replacing $\beta$ with $\beta F(\phi/\chi)$, some function of $\phi/\chi=\tanh\frac{\varphi}{\sqrt{6}\Mpl}$ which uniquely preserves the local conformal symmetry. 
The $\SO(1,1)$ symmetry is beneficial because it is restored in the large field limit $\varphi\to\pm\infty$. 
On the other hand, its explicit breaking naturally introduces the end of inflation.
Assuming $F(x)$ is odd, $P(X)$ has a non-trivial minimum $X=\frac{18\beta F\Mpl^4}{\gamma^4}$ for $F>0$ corresponding to the de Sitter phase, while such a minimum is not realized for $F<0$ because $X=\dot{\varphi}^2/2$ must be non-negative.
Thus, the graceful exit can be dynamically realized around the symmetry breaking point $\varphi \sim 0$.

\paragraph{Minimal slow-roll $k$-inflation.---\hspace{-0.8em}}

Thanks to the local conformal and broken global $\SO(1,1)$ symmetries, a flipping kinetic term of a hyperbolic-tangent form is naturally introduced.\footnote{The graceful exit making use of the hyperbolic-tangent switching of $X$ term itself is mentioned in Ref.~\cite{Kobayashi:2010cm} in the context of G-inflation and its gravitational reheating is discussed in Ref.~\cite{Hashiba:2018iff}.}
Such a Lagrangian is summarized as
\bae{\label{eq: Lagrangian}   
    \calL
    =\sqrt{-g}\left[\frac{\Mpl^2}{2}R-K(\Phi)X(\Phi)+\frac{1}{\Mpl^4}X^2(\Phi)\right], \nonumber \\
    K(\Phi)
    =\frac{6\beta}{\gamma^2}F\pqty{\tanh\pqty{\frac{\Phi}{\gamma\Mpl}}},
}
by field redefinition $\Phi\coloneqq\frac{\gamma}{\sqrt{6}}\VP$. Now the gravitational sector is reduced to the ordinary Einstein-Hilbert one and there remains no difference between the Riemannian and metric-affine approaches.
Thus it is a minimal realization of the original slow-roll $k$-inflation model~\cite{ArmendarizPicon:1999rj,Garriga:1999vw} with a hyperbolic-tangent kinetic term. Here we investigate its phenomenology as inflation, following Ref.~\cite{ArmendarizPicon:1999rj}.

At the background level, inflaton's pressure and energy density, $P$ and $\rho$, are given by
\bae{
    P=-K(\Phi)X
    +\frac{1}{\Mpl^4}X^2
    \qc \rho=2XP_X-P,
}
where $P_X=\partial P / \partial X$. The master equation of evolution is obtained as
\bae{\label{eq: master}
    \dot{\rho}=-\frac{\sqrt{3}}{\Mpl}\sqrt{\rho}(\rho+P)\propto P_X.
}
If $K(\Phi)$ is positive and its $\Phi$-dependence can be neglected, 
as we have mentioned, we have a stationary solution $\bar{P}_X\coloneqq\eval{P_X}_{X=\bar{X}}=0$
with $\bar{X}=\dot{\bar{\Phi}}^2/2=K\Mpl^4/2$ which gives the de Sitter universe ($\dot{\rho} = 0$). Here and hereafter barred quantities denote this de Sitter limit solution.

In realistic dynamics, $X$ however deviates from $\bar{X}$ due to the $\Phi$-dependence of $K(\Phi)$.
Corresponding to the slow-roll expansion in ordinary inflation models, one can expand the dynamics with respect to this deviation $\delta X\coloneqq X-\bar{X}$~\cite{ArmendarizPicon:1999rj,Garriga:1999vw}.
In fact, the master equation~\eqref{eq: master} indicates that the time dependence of the energy density dictating the slow-roll parameter $\epsilon=-\dot{H}/H^2$ is an $\calO(\delta X)$ quantity as
\bae{
    \!\dot{\rho}=-\frac{\sqrt{3}}{\Mpl}\eval{\sqrt{\rho}(\rho+P)}_{X=\bar{X}+\delta X}\!\simeq-\sqrt{3}K^2\Mpl\delta X,
}
at linear order in $\delta X$.
On the other hand, one can express it in terms of $K$'s time derivative via the leading form $\bar{\rho}=K^2\Mpl^4/4$ as
\bae{
    \dot{\rho}\simeq\frac{KK_\Phi\dot{\bar{\Phi}}\Mpl^4}{2},
}
where $K_\Phi = \partial K / \partial \Phi$.
Hence, as expected, $K$'s time derivative representing the deviation from the exact de Sitter universe is also an $\calO(\delta X)$ quantity as
\bae{
    \delta X\simeq-\sign(\dot{\bar{\Phi}})\frac{K_\Phi\Mpl^5}{2\sqrt{3K}}.
}
In order not to violate the null energy condition, i.e., $\rho + P > 0$, $\delta X$ should be positive. Once $\delta X$ is expressed in $K$'s derivatives, slow-roll parameters $\epsilon$, $\eta$, and $s$, and the sound speed $\cs$ can be obtained as
\bae{\label{eq: SR params}
    \bce{
        \dps
        \epsilon &\!\!\!
        \dps
        \coloneqq-\frac{\dot{H}}{H^2}\simeq-\sign(\dot{\bar{\Phi}})\frac{6\Mpl}{\sqrt{3}}\frac{K_\Phi}{K^{3/2}}, \\[5pt]
        \dps
        \eta &\!\!\!
        \dps
        \coloneqq
        \frac{\dot{\epsilon}}{H\epsilon}\simeq \sign(\dot{\bar{\Phi}})2\sqrt{3}\Mpl\left(\frac{
        K_{\Phi\Phi}}{K^{1/2}
        K_\Phi}-\frac{3}{2}\frac{
        K_\Phi}{K^{3/2}}\right),
        \\[5pt]
        \dps
        \cs^2 &\!\!\!
        \dps
        \coloneqq \frac{P_X}{P_X+2XP_{XX}}\simeq
        \frac{1}{12}
        \epsilon,
        \\[10pt]
        \dps
        s &\!\!\!
        \dps
        \coloneqq
        \frac{\dot{c}_\us}{H\cs}\simeq\frac{1}{2}\eta.
    }
}
The primary cosmological observables, the spectral index $\ns$ and the tensor-to-scalar ratio $r$, are expressed with use of these parameters as~\cite{Garriga:1999vw,Chen:2006nt}
\bae{
    \ns-1&\coloneqq\dv{\log\calP_\zeta}{\log k}\simeq 
    -2\epsilon-\eta-s
    \qc
    r
    \coloneqq\frac{\calP_h}{\calP_\zeta}\simeq
    16
    \cs\epsilon,
}
where $\calP_\zeta$ and $\calP_h$ denote the dimensionless power spectra of curvature and tensor perturbations, respectively. In addition, in such a kinetically driven inflation, $c_s \ll 1$ can be achieved and it predicts the large non-Gaussianity
which can be observed as the so-called equilateral shape in the bispectrum of the primordial curvature perturbations: $f_{\rm NL}^{\rm eq} \sim 1/\cs^2$ (see, e.g., Ref.~\cite{Chen:2006nt}).
Through the identity $\dd\Phi/\dd N=-\dot{\Phi}/H$, the backward e-folds $N$ from the end of inflation can be also obtained as
\bae{\label{eq: e-folds}
    N\coloneqq \log \left(\frac{a_{\mathrm{end}}}{a}\right)
    \simeq\frac{-\sign(\dot{\bar{\Phi}})}{2\sqrt{3}\Mpl}\int^{\bar{\Phi}}_{\bar{\Phi}_{\mathrm{end}}} \sqrt{K}\dd{\Phi},
}
where $\bar{\Phi}_{\mathrm{end}}$ denotes the field value at the end of inflation, i.e., $\epsilon=1$.

\paragraph{Large $\beta$ limit.---\hspace{-0.8em}} 

As we want to see the deviation from the exact de Sitter universe in this model, let us focus on the small field dynamics $\Phi\ll\gamma\Mpl$ where $\tanh(\Phi/\gamma\Mpl)$ can be approximated by $\Phi/\gamma\Mpl$.
This is actually accomplished in the large $\beta$ limit and we will see that the corresponding inflationary predictions are currently consistent with and testable in the future by cosmological observations.

Adopting the simplest form $F(x)=x$ for the symmetry breaking part, we hence approximate $K(\Phi)$ as
\bae{\label{eq: K as Phi}
    K(\Phi)\sim\frac{6\beta}{\gamma^3}\frac{\Phi}{\Mpl}.
}
With this simple assumption, the backward e-folds~\eqref{eq: e-folds} and slow-roll parameters~\eqref{eq: SR params} are easily obtained as
\bae{\label{field-N}
    \quad\frac{\bar{\Phi}}{\gamma\Mpl}=\pqty{\frac{9N^2}{2\beta}}^{1/3},
}
and
\bae{\label{eq: ns,r,cs}
    \ns-1
    \simeq-\frac{17}{6N} \qc
    r
    \simeq\frac{16\sqrt{2}}{9}\frac{1}{N^{3/2}} \qc
    \cs
    \simeq\sqrt{\frac{1}{18N}}.
}
Here we ignore $\bar{\Phi}_\mathrm{end}$.
The relation~(\ref{field-N}) in fact indicates that the small field assumption $\Phi\ll\gamma\Mpl$ is equivalent to the condition on $\beta$ as $\beta\gg\frac{9}{2}N^2$.
Note that the remained parameter $\gamma$ appears as a combination $\beta/\gamma^3$
in the power spectrum of the curvature perturbation given by
\bae{
     \calP_\zeta=\frac{1}{2\cs\epsilon\Mpl^2}\pqty{\frac{H}{2\pi}}^2\simeq\frac{81\cdot3^{1/3}}{16\cdot2^{1/6}\pi^2}\left(\frac{\beta}{\gamma^3}\right)^{4/3}N^{17/6}.
}
Thus, for each value of $\beta$, $\gamma$ should be fixed to be consistent with the CMB observation $\calP_\zeta\simeq2.1\times10^{-9}$~\cite{Akrami:2018odb}.

The energy and pressure density after the end of inflation in our model can be well approximated as $\rho\simeq P\simeq -KX$ since the square of the kinetic term becomes negligibly small, and the effective equation-of-state parameter becomes $w\coloneqq P/\rho\simeq1$, which phase is called \emph{kination}.
The energy density and the Hubble scale in the kination epoch evolve as $\rho\propto a^{-6}$ and $H^{-1}\propto a^{3}$ respectively, whereas these evolve as $\rho\propto a^{-3}$ and $H^{-1}\propto a^{3/2}$ in the ordinary inflaton oscillation phase (matter-domination).
Therefore the correspondence between the perturbation scale $k_*$ and the backward e-folds from the end of inflation is altered from the ordinary case, and assuming the instant transition at each phase of the universe, one finds the following formula
\bae{
    N=66-\ln\pqty{\frac{k_*}{0.002\,\mathrm{Mpc^{-1}}}}-\frac{1}{3}\ln\pqty{\frac{T_\uR}{10^6\,\mathrm{GeV}}} \nonumber \\
    -\frac{1}{6}\ln\pqty{\frac{g_*(T_\uR)}{106.75}}
    +\frac{1}{3}\ln\pqty{\frac{H_*}{H_\uf}}+\frac{1}{3}\ln\pqty{\frac{r_*}{0.005}}.
}
Here the reheating temperature $T_\uR$ denotes the radiation temperature at the time when the radiation energy density becomes comparable to the inflaton's one. $g_*$ stands for the effective degrees of freedom for energy density and we approximate it to be almost equal to ones for entropy density
at $T=T_\uR$. $H_*$ and $r_*$ are the Hubble parameter and the tensor-to-scalar ratio corresponding to the considered scale $k_*$, respectively.
$H_\uf$ is the Hubble parameter at the end of inflation.
We adopt the standard values for $g_*$ at the matter-radiation equality and the current matter density parameter $\Omega_\um$ as $g_*(T_\eq)=3.38$ and $\Omega_\um h^2=0.143$ with the normalized Hubble constant $h=H_0/(100\,\mathrm{km/s/Mpc})$, respectively~\cite{Aghanim:2018eyx}.

\begin{figure*}
    \centering
    \includegraphics[width=0.95\hsize]{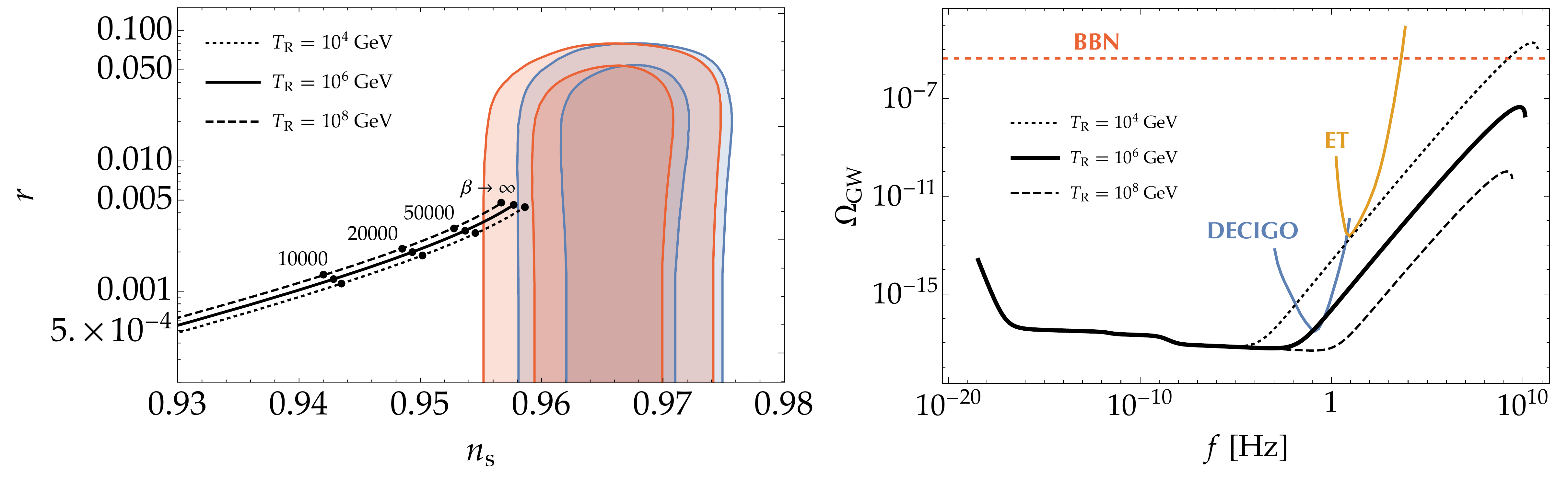}
    \caption{\emph{Left}:
    predictions for the spectral index $\ns$ and the tensor-to-scalar ratio $r$ with varying the kinetic coefficient $\beta$ and reheating temperature $T_\uR$.
    The black dotted, thick, and dashed lines correspond with $T_\uR=10^4\,\mathrm{GeV}$, $10^6\,\mathrm{GeV}$, and $10^8\,\mathrm{GeV}$, respectively, while the black points represent $\beta$'s value of $10000$, $20000$, $50000$, and the large $\beta$ limit~\eqref{eq: ns,r,cs} from left to right.
    We also plot the $1\sigma$ and $2\sigma$ constraints on the pivot scale $k=0.002\,\mathrm{Mpc}^{-1}$ from Planck TT, TE, EE+lowE+lensing+BK15 (red lines) and Planck TT, TE, EE+lowE+lensing+BK15+BAO (blue lines)~\cite{Akrami:2018odb}.
    \emph{Right}:
    the current density parameter of stochastic gravitational waves with respect to the reheating temperature $T_\uR$.
    Similarly to the left panel, the black dotted, thick, and dashed lines correspond with $T_\uR=10^4\,\mathrm{GeV}$, $10^6\,\mathrm{GeV}$, and $10^8\,\mathrm{GeV}$. The red dotted line indicates the BBN upper bound~\cite{Caprini:2018mtu}, while the blue and orange lines exhibit the future sensitivity prospects by DECIGO and Einstein Telescope observational projects (taken from Ref.~\cite{DEramo:2019tit}).
    }
    \label{fig: nsr}
\end{figure*}

The left panel of Fig.~\ref{fig: nsr} shows our numerical predictions on $n_\us$ and $r$ without the linear approximation of $\tanh(\Phi/\gamma\Mpl)$ in terms of the reheating temperature, varying $\beta$ (and $\gamma$ for a proper $\calP_\zeta$).
In the large limit of $\beta$ which well satisfies a condition $\beta\gg\frac{9}{2}N^2$, one confirms that predictions converge to the analytic formula~\eqref{eq: ns,r,cs}.
Predictions with large $\beta$ can be consistent with the Planck 2018's $2\sigma$ constraints~\cite{Akrami:2018odb,Akrami:2019izv}.
In particular, the tensor-to-scalar ratio $r \sim 0.005$ is sizable enough for a future test by CMB observations such as LiteBIRD~\cite{Hazumi:2019lys} and CMB-S4~\cite{Abazajian:2016yjj,Abazajian:2020dmr}.
One also finds, from the analytic formula~\eqref{eq: ns,r,cs}, the predicted sound speed $\cs \sim 0.03$ is consistently large with the $2\sigma$ allowed value $\cs\geq0.021$~\cite{Akrami:2019izv} from the non-Gaussianity on CMB, but characteristically small enough to be tested by, e.g., future galaxy observations~\cite{Yamauchi:2016wuc}.

\paragraph{Reheating and gravitational waves.---\hspace{-0.8em}}

As can be seen in the left panel of Fig.~\ref{fig: nsr},
the lower reheating temperature makes the $(\ns,\,r)$ prediction more consistent with the CMB observations.
As a possible reheating scenario in the kination phase, the so-called 
\emph{gravitational reheating}~\cite{Ford:1986sy,Starobinsky:1994bd} has been widely discussed, in which a part of the inflaton's energy is transferred into the radiation through the sudden change of the universe's expansion law between the inflation and kination eras.
Contrary to the ordinary reheating scenario~\cite{Kofman:1997yn}, the inflaton need not completely decay into the radiation because the inflaton's energy decreases as $\rho_\varphi\propto a^{-6}$, i.e., faster than radiational one $\rho_\ur\propto a^{-4}$.
Thus the radiation can dominate the universe sooner or later.
Consequently, the reheating temperature tends to be low in this scenario, and it is roughly given by $T_\uR\sim H_\uf^2/\Mpl$ (see, e.g., Ref.~\cite{Chun:2009yu}).
In our model, it corresponds with $T_\uR\sim10^6\,\mathrm{GeV}$ and we treat it as a fiducial value, though we keep allowing lower or higher reheating temperature as a free parameter
because it may depend on what kind of particles were produced and also how the transition from the inflationary phase to the kination proceeds~\cite{Kunimitsu:2012xx,Hashiba:2018iff}.
It also includes the other possibilities of reheating mechanisms~\cite{Tashiro:2003qp}.\footnote{Our inflaton is a pseudo Nambu-Goldstone boson of the $\SO(1,1)$ symmetry, so basically it only has derivative couplings to matter fields. The efficiency of reheating is hence expected not to be large anyway.}

The reheating temperature could be revealed by the stochastic gravitational waves~\cite{Nakayama:2008wy}. Some random gravitational waves are produced as tensor perturbations on superhorizon scales during inflation as well as scalar perturbations. After their horizon reentry, they behave as freely propagating radiation, and their relative energy density compared to the background inflaton can then grow as $\rho_\GW/\rho_\varphi\propto a^2$ during the kination era.
Therefore, for a lower reheating temperature or a longer kination phase, our inflation model predicts a larger stochastic gravitational wave amplitude~\cite{Giovannini:1999bh,Giovannini:1999qj,Babusci:1999ky,Riazuelo:2000fc,Tashiro:2003qp,Sami:2004xk,Artymowski:2017pua}.
In the right panel of Fig.~\ref{fig: nsr}, we show the prediction on the current density parameter $\Omega_\GW$ in terms of $T_\uR$ as well as sensitivity prospects of DECIGO~\cite{Seto:2001qf,Kawamura:2011zz} and Einstein Telescope projects~\cite{Punturo:2010zz}.
We also note that the gravitational wave amplitude is constrained from above not to spoil the successful Big Bang nucleosynthesis (BBN)~\cite{Cyburt:2015mya,Caprini:2018mtu}, whose bound is exhibited by the red dotted line.
One finds, for $10^5\,\mathrm{GeV}\lesssim T_\uR\lesssim10^6\,\mathrm{GeV}$, the predicted gravitational waves can evade the BBN constraint and can be tested by DECIGO.

\paragraph{Conclusions.---\hspace{-0.8em}}

In this Letter, we investigated a minimal realization of slow-roll $k$-inflation~\eqref{eq: Lagrangian} arose from the Lagrangian up to quadratic kinetic terms with the slightly broken global $\SO(1,1)$ symmetry in the conformal metric-affine geometry.
The left panel of Fig.~\ref{fig: nsr} shows the numerical predictions of the spectral index $\ns$ and the tensor-to-scalar ratio $r$ with varying the coefficient $\beta$. We found that the observational predictions on the spectral index $\ns$, the tensor-to-scalar ratio $r$, and the sound speed $\cs$, converge in the large $\beta$ limit (i.e. $K(\Phi)\sim 6\beta\Phi/(\gamma^3\Mpl)$ as shown in Eq.~\eqref{eq: K as Phi}) and become consistent with current observational data, particularly for lower reheating temperature. The intriguing aspect of our model is its testability by the forthcoming advancement of cosmological observations.
The predicted tensor-to-scalar ratio $r \sim 0.005$ is sizable enough to be checked by future CMB observation plans such as LiteBIRD~\cite{Hazumi:2019lys} and CMB-S4~\cite{Abazajian:2016yjj,Abazajian:2020dmr}.
In addition to the tensor-to-scalar ratio, the relatively small sound speed $\cs\sim 0.03$ will leave a characteristic non-Gaussianity on primordial perturbations, which is particularly beneficial to distinguish among inflationary models. The detailed study on the non-Gaussianity by using galaxies~\cite{Yamauchi:2016wuc} can potentially test our model as constraints on the sound speed get tighter. Furthermore, our model has also a prospect to be tested by future space-based gravitational wave detectors represented by DECIGO as depicted in the right panel of Fig.~\ref{fig: nsr} since the kination epoch intensifies the amplitude of stochastic gravitational waves at high frequency.

\acknowledgments

Y.T. is supported by JSPS KAKENHI Grants 
No. JP18J01992 and No. JP19K14707.
S.Y. is supported by JSPS Grant-in-Aid for Scientific Research
(B) No. JP20H01932 and 
(C) No. JP20K03968.

\bibliography{submit-minimal-k}
\end{document}